\documentclass[journal, 12pt, draftclsnofoot, onecolumn]{IEEEtran}

\usepackage{cite}      

\usepackage{subfigure}
\usepackage{wrapfig}
\usepackage{graphicx}  

\usepackage{psfrag}    

\usepackage{subfigure} 

\usepackage{url}       

\usepackage{stfloats}  
\usepackage{epsfig}
\usepackage{amsmath}   
\usepackage{amsmath, amssymb}

\begin{document}

\title{\LARGE{Design and Implementation of Distributed Resource Management Mechanisms for Wireless Mesh Networks}}

\author{George Athanasiou, Leandros Tassiulas
\thanks{G. Athanasiou is with the Automatic Control Lab, School of Electrical Engineering, KTH Royal Institue of Technology, Sweden. E-mail: \ttfamily{georgioa@kth.se}}%
\thanks{L. Tassiulas is with the Department of Computers and Communications Engineering, University of Thessaly, Greece. E-mail: \texttt{leandros@uth.gr}}%

}
\maketitle

\begin{abstract}
The wireless mesh 
networks are usually consisted of a static wireless backbone. The
main wireless mesh routers serve the routing of the traffic, while the peripheral routers act as access points (APs). These APs are
associated with the stations (STAs) in the network. The STA association/handoff procedures are very important towards achieving balanced operation in 802.11-based wireless mesh networks. In this paper we design and implement 
a resource management scheme based on cooperative association, where the STAs can share
useful information in order to improve the performance of the
association/handoff procedures. The cooperative association
mechanism is inspired by the rapidly designed cooperative protocols
in the field of wireless networks. Furthermore, we
introduce a load balancing mechanism that operates in a cross-layer manner taking
into account uplink and downlink channel conditions, routing performance and congestion control. The
iterative heuristic algorithms that we propose, control
the communication load of each mesh AP in a distributed manner. We
evaluate the performance of our mechanisms through OPNET
simulations and testbed experiments. 
\end{abstract}


%
\IEEEpeerreviewmaketitle


\section{Introduction}
The IEEE 802.11 \cite{80211} wireless local area networks (WLANs)
were originally designed to give a solution to the huge problem of
tangled cables of the end user devices. The stations (STAs) are
wirelessly connected to the available access points (APs) and the
APs are connected to a wired backbone network. The evolution of
these networks are the mesh networks where a wireless backbone
network is set up in order to support end-to-end wireless user
communication \cite{Akyildiz}. Several wireless routers that are
part of this wireless backbone network, forward the traffic in the
network. In addition a number of these routers also serve as APs.
The STAs are associated with the available APs and send their data
through them. Undoubtedly, the mesh networks are quite similar to
the infrastructure WLANs but they take an important benefit of their
self-organized structure and their dynamic nature. We can allegate
that the mesh networks are ad-hoc networks that operate in an
infrastructure mode. In this way they combine the benefits of ad-hoc
networks and WLANs.

In this paper we design and implement a cooperative association concept that
speeds up the basic association procedure. It is independent from
the association protocol that is active in the network. The main
outcome of this mechanism is that it eliminates the delays due to
scanning/probe and reassociation phases. The algorithm is called
"cooperative", since nodes share their view of the network with each
other. We have introduced a table that is maintained by each STA and
contains this information. This table contains the operational
frequencies of the mesh APs and their communication loads.
Therefore, each STA can make use of this information in order to
optimize its' association procedure. To the best of our knowledge
our cooperative mechanism is the first in this research field and is
fully compliant to 802.11.

Furthermore, in this work we present a novel load balancing scheme.
Wireless mesh networks can usually be overloaded due to the large
number of the associated STAs that try to send their data or due to
the large communication traffic that must be routed in the mesh
backbone. Our mechanism tries to overcome these situations by
controlling the communication load in a distributed manner. The APs
measure the load conditions in their neighborhood and inform their
associated STAs. The STAs execute a heuristic algorithm in order to
optimize their communication. Our load balancing mechanism uses
cross-layer information in order to provide a balanced MAC and
Routing layer operation.

The rest of the paper is organized as follows. In section II we
present the state of the art and briefly describe the main
restrictions in "expanding" the network capabilities that are
introduced by the current 802.11 association scheme. Section III
presents the application of a cross-layer association scheme in
wireless mesh networks that was proposed in \cite{Athanasiou}. In
section IV, we consider the cooperative association mechanism. In
section V, we describe a load balancing mechanism that guarantee a
jointly MAC/Routing balanced operation. Section VI presents the
simulation-based evaluation of the proposed mechanisms and section VII presents 
the implementation and the experimental evaluation of our work. Finally, in section VIII we
conclude and we pave the way for our future research directions.


\section{Association and Load Balancing schemes}
IEEE 802.11 defines an association procedure based on the RSSRI
(Received Signal Strength Report Indicator). The unassociated STAs
or the STAs that are trying to reassociate with a new AP, initialize
a scanning process to find the available APs that are placed nearby.
They measure the RSSRI values of each AP and associate with the AP
that has the highest RSSRI value (the strongest received signal).
Several studies have proved that the RSSRI-based association
mechanism can lead to poor network performance while the networks
resources are not utilized efficiently \cite{Arbaugh},
\cite{Bejerano1}. Therefore the research community focused their
research interests in designing new association methodologies that
will provide better resource utilization in the network. In
\cite{Athanasiou} the authors propose new dynamic association and
reassociation procedures that use the notion of the airtime cost in
making association decisions. In \cite{Bejerano2}, the authors study a
new STA association policy that guarantees network-wide max-min fair
bandwidth allocation in the network. In \cite{Korakis}, the authors
propose an association scheme that takes into account the channel
conditions (the channel information is implicitly provided by
802.11h \cite{80211h} specifications). The work in \cite{Kauffmann}
proposes an improved client association and a fair resource sharing
policy in 802.11 wireless networks. The authors in \cite{Arbaugh},
propose a technique to eliminate the probe phase delay of the
association process. The system presented in \cite{Bejerano1}
ensures fairness and QoS provisioning in WLANs with multiple APs.
Management frame synchronization is the basic part in the proposed
mechanism presented in \cite{Ramani}. In \cite{Shin}, the authors
formulate the association problem using neighbor and non-overlap
graphs. In \cite{Brik}, multiple radios are used in order to
implement more effective handoff mechanisms. In \cite{Kumar} the
problem of optimal user association to the available APs is
formulated as a utility maximization problem. The work in
\cite{Shakkottai} proposes a new mechanism where the traffic is
split among the available APs in the network and the throughput is
maximized by constructing a fluid model of user population that is
multi-homed by the available APs in the network. Finally, in \cite{Pack} there is an interesting study of
different fast handoff mechanisms.

Recently, there has been some work investigating the load balancing
problem in WLANs. In \cite{Garcia} the authors discuss some load
balancing approaches and how the future 802.11k \cite{80211k}
functionalities could contribute in implementing these mechanisms in
802.11 WLANs. The work in \cite{Sang} proposes a cross-layer
framework where packet level scheduling, handoff and system-level
load balancing are jointly performed. The techniques presented in
\cite{Bejerano3} control in an intelligent manner the cell size of
congested APs (cell-breathing), achieving in that way a load
balanced network operation. Last but not least, the work in
\cite{Bahl} proposes a new cell-breathing scheme that act as a load
balancing mechanism to control the congested cells in a WLAN.

The work in \cite{Athanasiou} introduces a cross-layer association
framework that works in the direction of optimizing the association
decisions of the STAs in the network. A STA that is initiating a reassociation process
has to scan for available APs in its' neighborhood. This scanning is
a time consuming process and slows down the handover. In order to
eliminate these delays we propose a new association concept where we
support STA collaboration. Furthermore, in the cross-layer
mechanisms we have observed especially in high load conditions that
the mesh APs that provide good QoS are usually overloaded.
Communication overload is quite often in pure 802.11 operation too,
even in low load conditions. Consequently, the proposed load
balancing scheme tries to cope with this problem and provide a
balanced network operation.

\section{Applying Cross-Layer Association in 802.11-based Mesh Networks}
In this section we provide a high level overview of the cross-layer
association mechanism that is proposed in \cite{Athanasiou} which is
the basis of the proposed mechanisms in this paper.

The proposed end-to-end QoS aware cross-layer association scheme is
designed for 802.11-based mesh networks. In this sophisticated
mechanism the association decision is based on a new metric called
airtime cost that reflects the average duration for which the
channel is occupied. The authors define the airtime cost of station
\(i \in U_a\), where \(U_a\) is the set of stations associated with
AP \(a\), as:
\begin{equation}
C_a ^i  = \left[ {O_{ca}  + O_p  + \frac{{B_t }}{{r^i }}}
\right]\frac{1}{{1 - e^i _{pt} }}.\label{eq_airtime}
\end{equation}
In \eqref{eq_airtime}, \( O_{ca}\) is the channel access overhead,
\(O_p\) is the protocol overhead and \(B_t\) is the number of bits
in the test frame. Some representative values (in 802.11b networks)
for these constants are: \(O_{ca}=335\mu\)s, \(O_p=364\mu\)s and
\(B_t=8224\)bits. The input parameters $r^i$ and \(e_{pt}\) are the
bit rate in \(Mbs^{-1}\), and the frame error rate for the test
frame size \(B_t\), respectively.

The load on the "uplink" channel of a particular AP $a$ is defined
as:
\begin{equation}
C_a^{up}  = \left[ {O_{ca}  + O_p  + B_t \overline
{({\raise0.7ex\hbox{$1$} \!\mathord{\left/
 {\vphantom {1 {r^{up} }}}\right.\kern-\nulldelimiterspace}
\!\lower0.7ex\hbox{${r^{up} }$}}} )} \right]\frac{1}{{1 - \overline
{e_{pt}^{up} } }}\left| {U_a } \right|,
\end{equation}
where \(\overline{e_{pt}^{up}}\), \(\overline {r^{up}}\), and
\(\left| {U_a } \right|\) are the average uplink error probability,
average uplink transmission rate and the number of STAs associated
with AP $a$ respectively. The load on the "downlink" is defined as:
\begin{equation}
C_a ^{down}  = [O_{ca}  + O_p ]\sum\limits_{j \in U_a } {\frac{1}{{1
- e_{pt}^j }}}  + B_t \sum\limits_{j \in U_a } {\frac{1}{{r^j (1 -
e^j _{pt} )}}}.
\end{equation}
The main idea of this mechanism is that the STAs base their
association decision on the cumulative uplink and downlink airtime
cost. Specifically, each AP $a$ broadcasts with its beacons,
information about \(\overline{e_{pt}^{up}}\), \(\overline {r^{up}}\), and
\(\left| {U_a } \right|\). The STA receives this information and
calculates \(C_a^{up}\). It also receives \(C_a^{down}\) from each
AP $a$ and finally selects for association the AP with the minimum
\(C_a^{up}+C_a^{down}\). In this way the STA can be associated with
the less loaded AP that can provide the best communication channel.

In order to apply this mechanism in 802.11-based wireless mesh
networks the authors extended the previous mechanism in a
cross-layer manner, having in mind the end-to-end QoS provisioning
in the network. In a wireless mesh network there is a popularity of
mesh APs that are available for association. In order to achieve
high QoS provisioning we must consider the end-to-end case where the
wireless backhaul network plays an important role.

The upcoming 802.11s \cite{80211s} standard introduces the airtime link cost as
the default routing metric in the mesh backhaule. The Radio
Metric-Ad Hoc On Demand Distance Vector (RM-AODV) protocol modifies
the basic AODV protocol by selecting the end-to-end path with the
minimum total airtime cost.

In this mechanism there is a combination of the routing airtime
costand the association airtime cost. The total end-to-end airtime
cost is defined as:
\begin{equation}
TC_i^{rcv} = (AC_i ^{up}  + AC_i^{down} )w_1  + RC_i^{rcv} w_2,
\end{equation}
where \(TC_i^{rcv}\) is the total weighted cost calculated for MAP
$i$, \(AC_i ^{up}, AC_i ^{down}\) are the association airtime costs
for the uplink and downlink respectively, \(RC_i^{rcv}\) is the
routing airtime cost for the path from MAP $i$ to the receiver
\(rcv\) and \(w_1, w_2\) are the weights.

The main idea in this end-to-end QoS aware mechanism is that the
STAs operate in an active way by sending probe frames to the
candidate for association APs. We can call these probe frames
"per-receiver association requests" that are used by the STAs
to announce their data receiver in order to come up with
optimal association decision.
The RM-AODV calculates the routing airtime cost from each AP to the
data receiver and these values are broadcasted by each
candidate AP. The STAs follow the previous procedure in parallel
with the calculation of the association airtime cost (using uplink and
downlink costs). Finally the association airtime cost and the
routing airtime cost are weighted. Each STA calculates
\(TC_i^{rcv}\) and selects for association the mesh AP with the
minimum total airtime cost. The previous mechanism can be
dynamically executed in a periodic manner by each STA in order to
initiate reassociation process in case that a new AP can provide
better QoS in the network. An interested reader can refer to
\cite{Athanasiou} for a detailed description of the protocol.

\section{Cooperative Association Mechanism} In this section, we
introduce a cooperative association mechanism whose goal is to speed
up the basic association process. This mechanism is independent from
the underlying association decision protocol. However, in this work
we jointly use the cooperative mechanism with the cross-layer
association mechanism in order to optimize the handoff delay, and
thus, the end-to-end delay.

In order to make clear the importance of the proposed protocol, we analyze the delay that is introduced by the basic 802.11-based association process \cite{Arbaugh}. The delay of the
802.11-based handover process can be divided into three categories:
\begin{itemize}
\item{\textbf{Probe or scanning delay}: During the first step in the association procedure that is
determined by 802.11 the STAs have to scan for available APs and
"hear" their beacon frames. This is a time consuming procedure since
the STAs must scan all the available channels (12 for 802.11a)  in
order to find active APs. Furthermore, the STAs have to follow the beacon intervals for
data synchronization reasons.  Scanning delay constitutes a major
portion of the handover delay.}
\item{\textbf{Authentication delay}: After the
scanning phase the STAs have to exchange authentication frames in
order to be authenticated by the current AP.}
\item{\textbf{Reassociation delay}:
When a STA moves from an AP to a new AP, it has to exchange
reassociation frames with the new AP. }
\end{itemize}

In a wireless mesh network the STAs can cooperate to share their
view of the network.  In particular, the STAs already associated
with the network can relay the information regarding different APs
in the network to the STAs not yet associated or the STAs
re-associating.  By this way, STAs will obtain information about the
available APs in an expedited fashion, effectively removing delays
due to scanning the channels.

\begin{figure}
\centering
\includegraphics[width=2.5in]{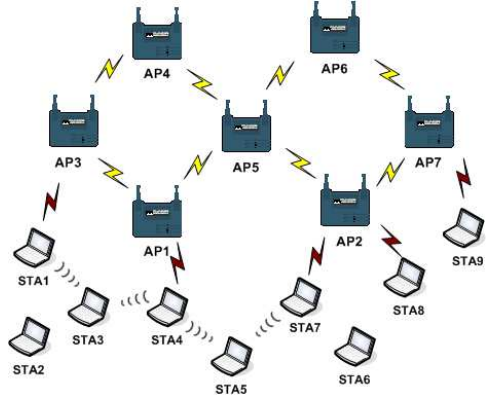}\vspace{-0.1in}
\caption{Cooperative association in a wireless mesh network.}
\label{fig1}\vspace{-0.2in}
\end{figure}

We discuss the basic idea of our cooperative mechanism through the
description of an example mesh network scenario. Figure \ref{fig1}
depicts a mesh network where STA1, STA4, STA7, STA8 and STA9 are
already associated with the available APs. In this figure we
consider a scenario where STA3 and STA5 have just turned on and are
initiating an association process in order to associate with an AP.
Traditionally these STAs must scan the available channels in order to find an AP for association. The proposed approach aims at
eliminating the scanning delay. This can be achieved by introducing
cooperation between the wireless STAs. If STA1 and STA4 informs STA3
about the operational frequencies of AP3 and AP1, STA3 does not need
to spend time in the scanning phase. Besides, STA1 and STA4 already
know the uplink and downlink channel conditions (in the
communication with the APs that are currently associated with) and
can inform STA3 about these. Consequently STA3 can be aware of the
operational frequency and the load of the AP3 and AP1.

The basic component in our cooperative mechanism is a special table
that is maintained by each associated STA. This table contains the
information about the channel and load conditions of the available
APs. It is called {\em ASSOC\_TABLE} and is regularly updated by the
STAs. {\em ASSOC\_TABLE} is depicted in table I:

\begin{table}
\caption{\em ASSOC\_TABLE structure}{
\begin{center}\vspace{-0.1in}
\begin{tabular}{|c|c|c|c|}
\hline
$\textbf{MAC}$&$\textbf{CHANNEL}$&$\textbf{LOAD}$&$\textbf{T/STAMP}$\\
\hline\hline
$MAC(AP1)$&$1$&$13$&$123$\\
\hline
$MAC(AP2)$&$6$&$45$&$134$\\
\hline
$MAC(AP3)$&$11$&$33$&$136$\\
\hline
\end{tabular}
\end{center}}\vspace{-0.4in}
\end{table}

In the {\em ASSOC\_TABLE} the load of the APs is represented by
their cumulative uplink and downlink airtime cost. We keep a separate timestamp for each table record in order to check the suitability of the stored information.

In the context of our cooperative mechanism the STAs can operate in
two modes. In the first mode STAs periodically broadcast their {\em
ASSOC\_TABLE} tables. By this way, the STAs that are in close
proximity to each other can keep up to date their channel and load
information of the APs in the network. In the second mode of
operation, a newcomer STA first sends a control frame to its
neighbors in order to acquire the appropriate information.  This
control frame is an enhanced "`Cooperative Probe Request"' frame also containing any
information related to the association. The neighboring STAs
receiving this control frame respond with the broadcast of their
{\em ASSOC\_TABLE}. The newcomer STA builds its own {\em
ASSOC\_TABLE} according to the information collected from its
neighbors. In our simulations, both of these modes are implemented.

\begin{figure}
\centering
\includegraphics[width=3.0in]{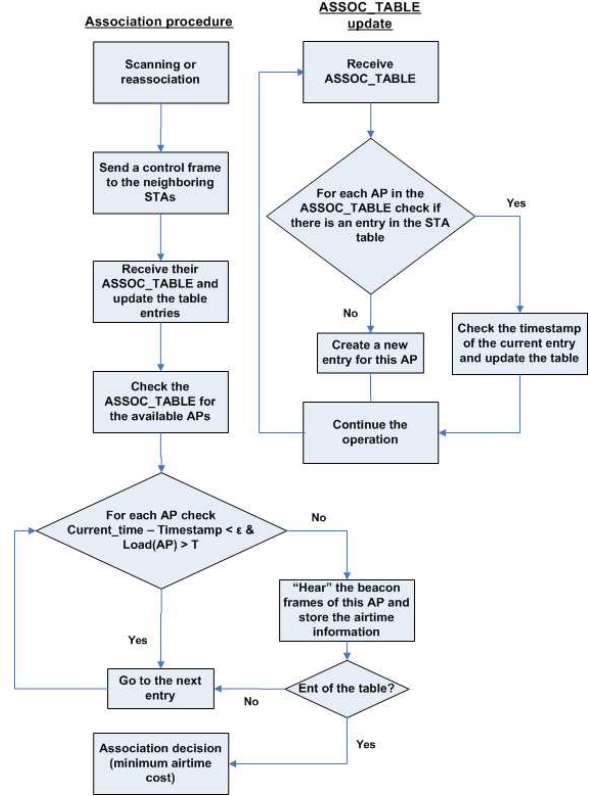}\vspace{-0.1in}
\caption{The main process executed by each STA.} \label{fig2}\vspace{-0.2in}
\end{figure}

Figure \ref{fig2} depicts the flow chart that is executed when a STA
receives the broadcasted {\em ASSOC\_TABLE}. Firstly the STA checks
if there is any entry in the table that contains information for the
current AP. In case that there is an entry, the STA updates the
corresponding information and otherwise, the STA generates a new
entry inside the table.

As we have mentioned, in order to avoid scanning delay, the STAs use the information stored in their
{\em ASSOC\_TABLE}. In case that a STA initiates a scanning or a
reassociation process it has to execute the procedure that is
depicted in Figure \ref{fig2}. As depicted in Figure \ref{fig2} for
each AP there are two separate checks. First, the suitability of the
stored information is checked by comparing the timestamp of the specific record
with the current time. In addition we check the load of the current
AP. We have introduced a threshold that helps us avoid the
overloaded situations in the network. The main idea of this
threshold is that in case that an AP is overloaded we must exclude
it from the list of the candidate APs for association.

In our proposed cooperative association mechanism, the STAs operate
in infrastructure and ad-hoc modes.  Infrastructure mode is used to
send data through the network and ad-hoc mode is used to communicate
with the neighboring STAs and to relay the {\em ASSOC\_TABLE}.
There are several ways to achieve this behavior in a STA. One is to have the wireless card dynamically changing the mode from infrastructure to ad-hoc, every time the STA needs to execute the cooperative association procedure. Being in this mode, the node can communicate with neighbors that operate in the same channel and exchange information with them about existing APs. An alternative would be the use of a "`control channel"' where all the stations would exchange scanning information. It is obvious that in the proposed scheme, the stations still have to execute the scanning procedure in order to update their {\em ASSOC\_TABLE}. Nevertheless, now the procedure is executed less often, since the stations receive association information from neighbors.


\section{Load Balancing Mechanism}
In this section we propose a novel load balancing scheme that is
based on the airtime information. First, we describe the main idea
of the load balancing scheme: In dynamic mesh networks there is a
variety of APs that are available for association. Traditionally the
RSSRI-based association mechanism measures the channel strength of
the available APs and leads the STAs to be associated with the APs
that provide the "clearest" or "strongest" channel. Unfortunately
this can lead to overloaded situations. In \cite{Athanasiou} the
authors proposed a sophisticated cross-layer association mechanism
that tries to phase this problem. Routing is an important component
of this association mechanism. The STAs obtain the appropriate
information for the uplink and downlink channels and the mesh
backhaul in order to optimize their association decision. In our
experiments, we have seen that in high load network conditions the
performance of the network decreases. There is a demand for a
sophisticated mechanism that can balance the load of the network.
This mechanism extends the performance bounds of the cross-layer
association mechanism by controlling the communication load in an
autonomous manner.

In our load balancing scheme the total airtime cost for the mesh AP
\(i\) in routing the data to the receiver \(rcv\) is defined in
equation (4). In \cite{Athanasiou} there is a detailed description
of the computation of the total airtime cost. The main idea is that
a STA optimizes its' association decision by computing the total
airtime costs in their neighbor and associating with the mesh AP
that can provide the best QoS (that has the lower total airtime
cost).
The weights \(w_1\) and \(w_2\) play an important role in the
computation of the total airtime cost of a particular data
transmission and therefore they determine the association decision
of a STA. In our load balancing mechanism we apply a heuristic
algorithm to adaptively change these weights "on the fly", according
to the communication load of the cells.

We now describe an example communication scenario which shows the
importance of a load balancing mechanism. As we mentioned before, an AP in the mesh network can guarantee data routing through efficient routes to a possible receiver (routes with low cumulative routing airtime cost). In case that there is a huge amount of neighboring STAs associated with this AP, it is possible to face a bottleneck at the first, wireless hop of the transmission. In order to
overcome these situations, we have to adaptively control the load of
the current cell. A cell breathing strategy based on the weight
adaptation is executed. By increasing w1 we enforce the impression
of the association airtime cost (that reflects the AP load) in the
total airtime cost. By this way the STAs that are associated with
the current AP will be aware of the communication overload. During
the following subsection we describe in detail this mechanism.

We define a balancing index \(b\) (introduced in \cite{Chiu}) that
reflects the load conditions in a neighbor \(n\) (\(n\) neighboring
APs) as:
\begin{equation}
b = \frac{{(\sum\limits_{i = 1}^n {AC_i } )^2 }}{{n\sum\limits_{i =
1}^n {AC_i ^2 } }}
\end{equation}
where \(AC_i=C_i^{up}+C_i^{down}\) for each AP \(i\). Each AP \(i\)
broadcasts its' \(b\) value to the neighboring APs in a periodic
manner. This broadcasting can be performed using the Local
Association Base Advertisement (LABA) mechanism that is proposed in
802.11s standard. This mechanism informs the entire mesh network
about the STAs associated with the available mesh APs. The mesh APs
periodically broadcast LABA messages that contain useful
information. We can extend these messages by incorporating in them
the airtime information of each AP. Consequently, the mesh APs are
aware of the channel communication conditions of each mesh AP in the
network.

Considering that we have two neighboring APs we compute the load
balancing index:
\begin{equation}
b = \frac{{(AC_1  + AC_2 )^2 }}{{2(AC_1^2  + AC_2^2 )}} = \left\{
{\begin{array}{*{20}l}
{1,\ AC_1=AC_2\ (balanced)}  \\
{\frac{1}{2},\ AC_1=0\ or\ AC_2=0}  \\
{1 > b > \frac{1}{2},\ otherwise}  \\
\end{array}} \right.
\end{equation}
Generally for $n$ neighboring APs:
\begin{equation}
b = \left\{ {\begin{array}{*{20}l}
{1,\ AC_i=AC_j=\ldots=AC_k\ (balanced)}  \\
{\frac{1}{n},\ AC_1=0\ or\ AC_2=0\ or \ldots\ AC_n=0}  \\
{1 > b > \frac{1}{n},\ otherwise}  \\
\end{array}} \right.
\end{equation}

\begin{figure}
\centering
\includegraphics[width=2.5in]{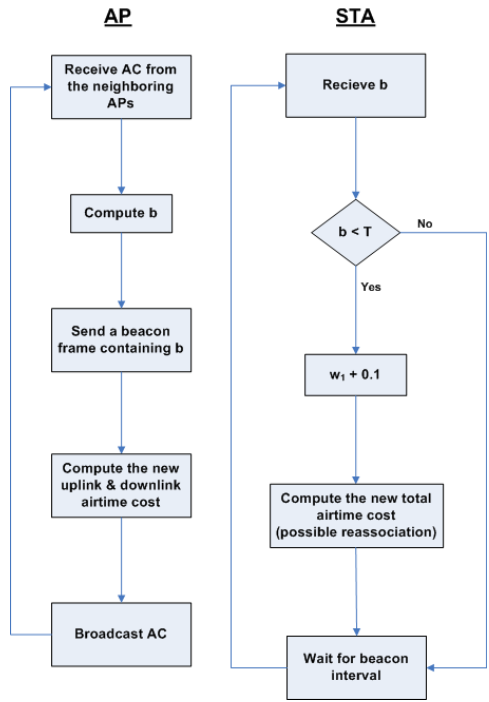}\vspace{-0.1in}
\caption{The heuristic load balancing algorithm} \label{fig4}\vspace{-0.2in}
\end{figure}

In a neighborhood that contains $n$ APs we introduce a threshold \(1
> T \ge {\raise0.7ex\hbox{$1$} \!\mathord{\left/
 {\vphantom {1 n}}\right.\kern-\nulldelimiterspace}
\!\lower0.7ex\hbox{$n$}} \) which represents the lower bound of a
balanced network operation. In case \(b < T \) the STAs must
increase weight \(w_1\) in order to make an association decision
based mainly on the association airtime cost (AC). The cumulative
association airtime cost represents the load of an AP. In our
experiments we have seen that in high loaded networks we must chose
a threshold close to 1 and in very low load conditions we must chose
a threshold close to \({\raise0.7ex\hbox{$1$} \!\mathord{\left/
 {\vphantom {1 n}}\right.\kern-\nulldelimiterspace}
\!\lower0.7ex\hbox{$n$}} \). We have run different load scenarios in
order to choose the appropriate \(T\) values. The diagrams in
figure \ref{fig4} depict the execution of the
heuristic algorithm that performs communication load balancing in
the mesh network based on the balancing index that we have described
before.

During the execution of our heuristic algorithm the mesh APs
periodically receive the airtime cost values from the neighboring
mesh APs and compute the balancing index $b$. The mesh APs broadcast
this index in the beacon frames and their associated STAs can "hear"
them, extract the carried information and compare $b$ with the
predefined threshold. In case \(b < T\) there is an unbalanced
network operation in this neighborhood, the STAs increase the value
of \(w_1\) and compute the new total airtime cost. At this step of the
heuristic algorithm STA reassociations are quite often due to the
modification of \(w_1\).


\section{Simulation-based Evaluation}
The proposed mechanisms have been implemented in OPNET\cite{OPNET}.
We have built our protocols on
top of the IEEE 802.11 standard and in this way we can achieve
backward compatibility. We have modified the main control frames
(beacon and probe frames) in order to incorporate the appropriate
information in them. The light modifications that we have introduced
in the basic functionality of the IEEE 802.11 standard do not affect
the performance of the network.


\subsection{Multi-Cell Scenario}
We first study a multi-cell 802.11 network that consists of four
overlapping cells. In such simple topologies we can control the
parameters of our system and we can have a clear view of the
performance of the proposed protocols. The STAs are uniformly
distributed in the network and their data frames are transmitted at
1024kbps. We compare the performance of the basic 802.11-based
association mechanism, the same mechanism equipped with cooperation
between the stations, the airtime association mechanism and the
airtime mechanism equipped with STA cooperation, while the
communication interference changes during the network operation.

We study the performance of the proposed mechanisms while the number
of the STAs in the network grows. During our simulation scenarios
the number of the associated STAs in the network increases from 5 to
65 (STAs are uniformly placed in the network). We measure the
network throughput, the average transmission delay and the data
dropping. We believe that these measurements are representative and
reflect the system performance under different operational
conditions.

\begin{figure} 
\centering
\includegraphics[width=3.0in]{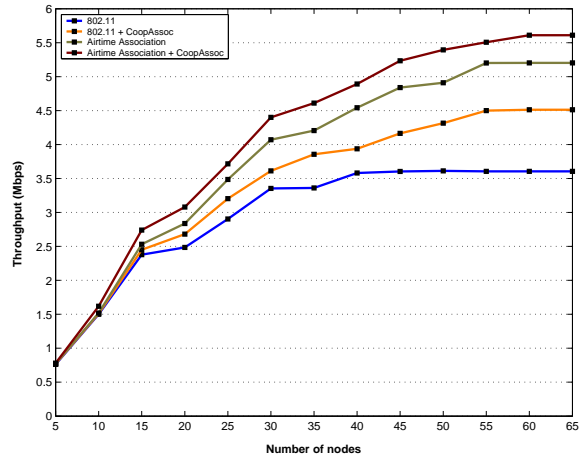}\vspace{-0.1in}
\caption{Throughput (multi-cell scenario).} \label{fig6}\vspace{-0.2in}
\end{figure}

Figure \ref{fig6} depicts the network throughput while the number of
the associated STAs increases. We compare the throughput values that
are achieved during the execution of the airtime association
mechanism, the cooperative association mechanism and the basic
802.11-based association scheme. It is clear that the highest
throughput values are achieved when we apply jointly the airtime
association and the cooperative association mechanism. Contrarily,
the 802.11 has the worst performance during our studies. We have to
point out that the cooperative mechanism that we have introduced in
this work speeds up the association procedure. We jointly apply this
mechanism with the airtime and the basic 802.11-based association
mechanism. The results are depicted in figure \ref{fig6}. In low
load conditions we observe a quite small throughput improvement by
the use of the proposed mechanisms. In high load conditions
throughput increase is higher. When we apply cooperative association
to the original 802.11 we can see an improvement of approximately
30\% in terms of throughput. The maximum throughput improvement that
is achieved by the combined application of the airtime and CoopAssoc
mechanism is approximately 55\% (when we have 65 associated STAs).
It is important to notice that the 802.11 network throughput is
stabilized when we have 45 associated STAs in the network. This
means that after that point the provided QoS in the network is
getting worse while the number of the STAs in the network increases.
Meanwhile, the airtime and CoopAssoc mechanisms expand the network
capabilities and maximize the network throughput in presence of 65
associated STAs in the network. This expansion is true due to the
sophistical operation of the airtime association mechanism (that
takes into account the uplink and downlink channel information) and
the fast operation of the CoopAssoc (that gets rid of the time
consuming association procedures).
\begin{figure}
\centering
\includegraphics[width=3.0in]{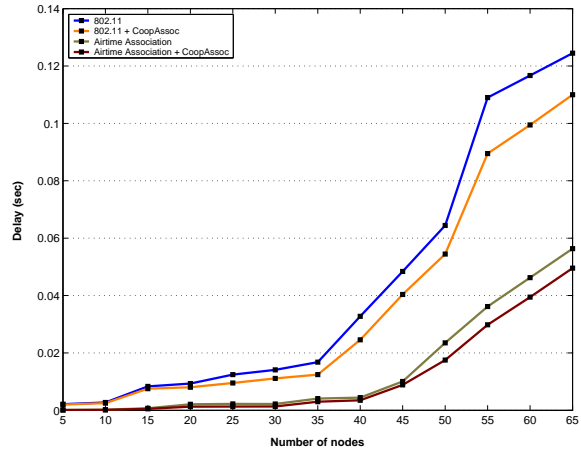}\vspace{-0.1in}
\caption{Average transmission delay (multi-cell scenario).}
\label{fig7}\vspace{-0.2in}
\end{figure}

In figure \ref{fig7} we observe the average transmission delay in
the network. It is clear that in low load network operation the
average transmission delay of the 802.11 is quite small and close to
the average delay that is achieved by the airtime and the CoopAssoc
mechanisms. While the number of the associated STAs increases over
35 the average delay of the 802.11 is getting extremely high. In
contrary the airtime and the CoopAssoc mechanisms keep the
transmission delay in low level. The 802.11-based association policy
is quite static and the main outcome of this characteristic is that
some cells are overloaded in the network while the number of the
associated STAs increases. In other words the 802.11 doesn't have
the capability to provide a balanced network operation in high load
conditions. The basic feature in the operation of the airtime and
the CoopAssoc mechanisms is that they are aware of the communication
load in the cells of the network. CoopAssoc provides fast dynamic
reassociations in order to overcome the overloaded problem and keep
a balanced network operation. The huge 802.11 scanning delays
are avoided as the CoopAssoc mechanism "grants" the appropriate
information to the STAs that are trying to be reassociated with a
new AP.

\begin{figure}
\centering
\includegraphics[width=3.0in]{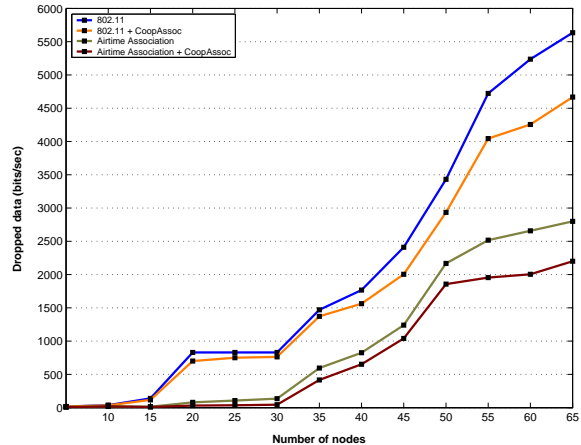}\vspace{-0.1in}
\caption{Average dropped data (multi-cell scenario).} \label{fig8}\vspace{-0.2in}
\end{figure}

Figure \ref{fig8} depicts the amount of data dropped due to channel
errors, contention and collisions. The airtime and the CoopAssoc
mechanisms achieve lower data dropping through the communication in
the network. The sophisticated channel-based association policies
that are introduced by the airtime mechanism and the fast
association/reassociation procedures that are introduced by the
CoopAssoc enforce the dynamic behavior of each STA in the network.
Practically this means that the STAs are aware of the communication
conditions in the cells of the network and can rapidly adapt their
behavior in order to achieve higher QoS provisioning.

\subsection{Mesh Network Scenario}
In order to measure the end-to-end network performance we study the
application of the proposed mechanisms in a 802.11-based wireless
mesh network. We have simulated a wireless mesh network in the OPNET
simulation environment. The wireless routers that are provided by
the OPNET wireless module are part of the backhaul network. The
peripheral routers serve as APs as well. In the cross-layer
mechanisms we need information from the routing layer. Therefore, we
have modified the basic AODV routing protocol protocol in order to
implement a link quality aware AODV (where the airtime cost serves
as an a routing decision metric). In other words we have implemented
RM-AODV that is introduced by 802.11s standard and we have applied
this routing protocol at the mesh backhaul. In addition, we have
implemented a cross-layer interface which supports the passing of
information from one layer to another while the OPNET suite is not
equipped with such mechanisms. The STAs are uniformly distributed in
the wireless mesh network. For the communication between the
wireless routers in the backhaul network, we use the physical model
of IEEE 802.11a OFDM physical layer. The supported physical rate of
this scheme is 12 Mbps. The STAs that are associated with the
available peripheral APs transmit their packets at 1024 kbps. We
have simulated the cross-layer association, the cooperative
association and the load balancing mechanisms by introducing light
modifications in the basic functionality of 802.11 (frame
modifications, etc.).

We first introduce FTP traffic in the network. In the OPNET
simulation environment an FTP server is directly connected to the
wireless backhaul network that supports file upload and download. In
our first experiment we continually increase the number of the STAs
in the network and we keep fixed the file size to 500 Kb. During the
second experiment we uniformly place 15 STAs in the network and we
fluctuate the files size that the STAs upload or download.

\begin{figure*}[t]
\centering
\parbox{1\textwidth}{
\subfigure[Throughput vs number of
stations.]{\parbox{0.51\textwidth}{
\includegraphics[width=3.0in]{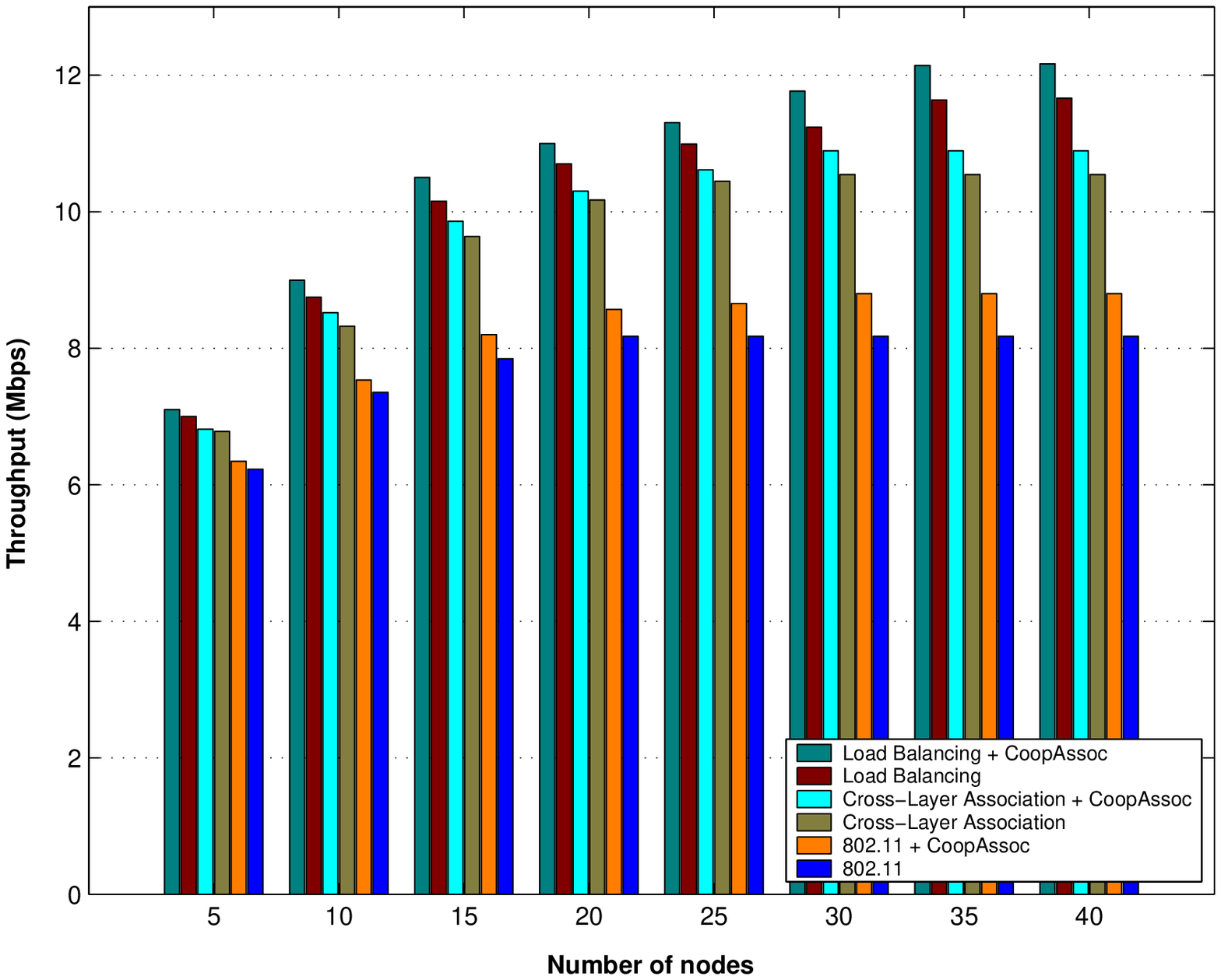}}\label{fig9a}}
\subfigure[Throughput vs file size.]{\parbox{0.51\textwidth}{
\includegraphics[width=3.0in]{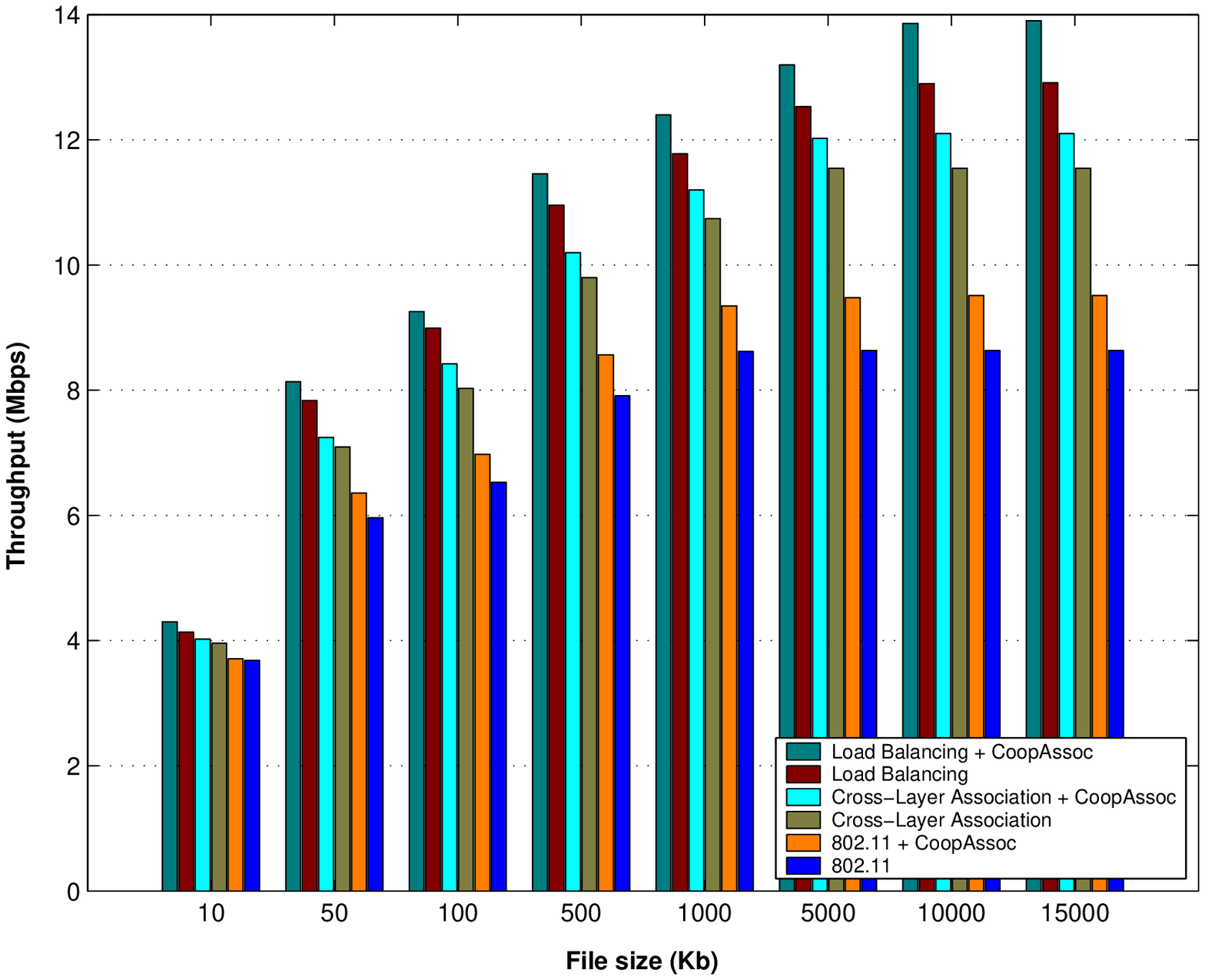}}\label{fig9b}}\vspace{-0.1in}
\caption{FTP simulation results.}} \label{fig:fig9}\vspace{-0.2in}
\end{figure*}

In figure \ref{fig9a} we observe the throughput mutation while the
number of the associated STAs in the network increases from 5 to 40.
We compare the performance of the proposed mechanisms that are
applied jointly in some cases with the basic 802.11-base association
scheme. As we have mentioned CoopAssoc is an independent mechanism
and therefore it can be applied jointly with the cross-layer
association or the load balancing mechanisms. The load balancing
mechanism enhanced with STA cooperation achieves the higher
throughput values in the network. A remarkable outcome of this
figure is that the 802.11 throughput is stabilized over the 20
associated STAs in the network, while the application of the
cross-layer association mechanism maximizes the network throughput
in presence of 30 STAs. In contrary, the load balancing mechanism
keeps increasing the network throughput and it maximizes it in
presence of 40 associated STAs. First of all, the incapability of
the RSSRI-based association scheme (in 802.11) to provide high
end-to-end QoS in a wireless mesh environment is quite observable.
The static association policies that are introduced by the
RSSRI-based mechanism encourage static associations between the STAs
and the APs. However the mesh networks are dynamic communication
environments where the channel conditions are time varying. The STAs
must be aware of the uplink and downlink channel characteristics in
order to adapt their behavior and enjoy higher QoS in the network.
The cross-layer association mechanism provides the channel and the
routing information from the backhaul to the STAs. The STAs can make
good use of this information to optimize their association decision.
In addition the load balancing mechanism uses the same information
to provide a "communication breathe" to the overloaded cells. The
cross-layer association mechanism (enhanced with cooperation)
achieves 35\% throughput improvement while the load balancing
mechanism (enhanced with cooperation) achieves approximately 49\%
throughput increase.

Figure \ref{fig9b} depicts the throughput variation while the size
of the uploaded/downloaded files increases from 10 Kb to 15000 Kb.
For the same reasons that we have mentioned before the load
balancing mechanism enhanced with cooperation attains the highest
throughput values in the network. The cross-layer association
mechanism (enhanced with cooperation) achieves approximately 43\%
throughput increase while the load balancing mechanism (enhanced
with cooperation) achieves 64\% throughput increase.

During the second experiment we have simulated a VoIP application in
the same 802.11-based wireless mesh network. In our simulations we
have uniformly placed several VoIP clients in the network. We run
different simulation scenarios where we vary the number of the VoIP
sessions that are supported in parallel.

\begin{figure*}[ht]
\centering
\parbox{1\textwidth}{
\subfigure[Average client access delay.]{\parbox{0.51\textwidth}{
\includegraphics[width=3.0in]{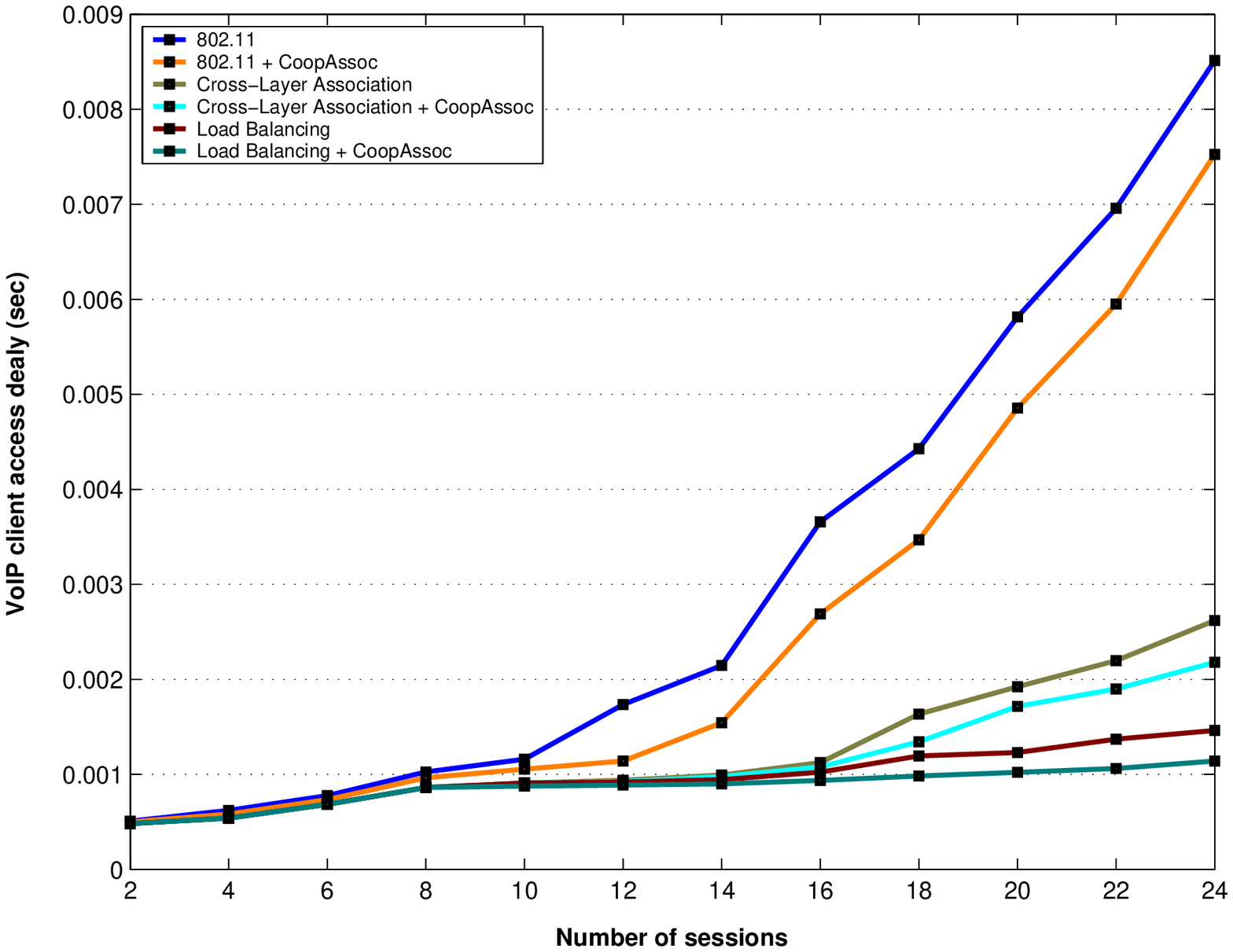}}\label{fig10a}}
\subfigure[Average AP access delay.]{\parbox{0.51\textwidth}{
\includegraphics[width=3.0in]{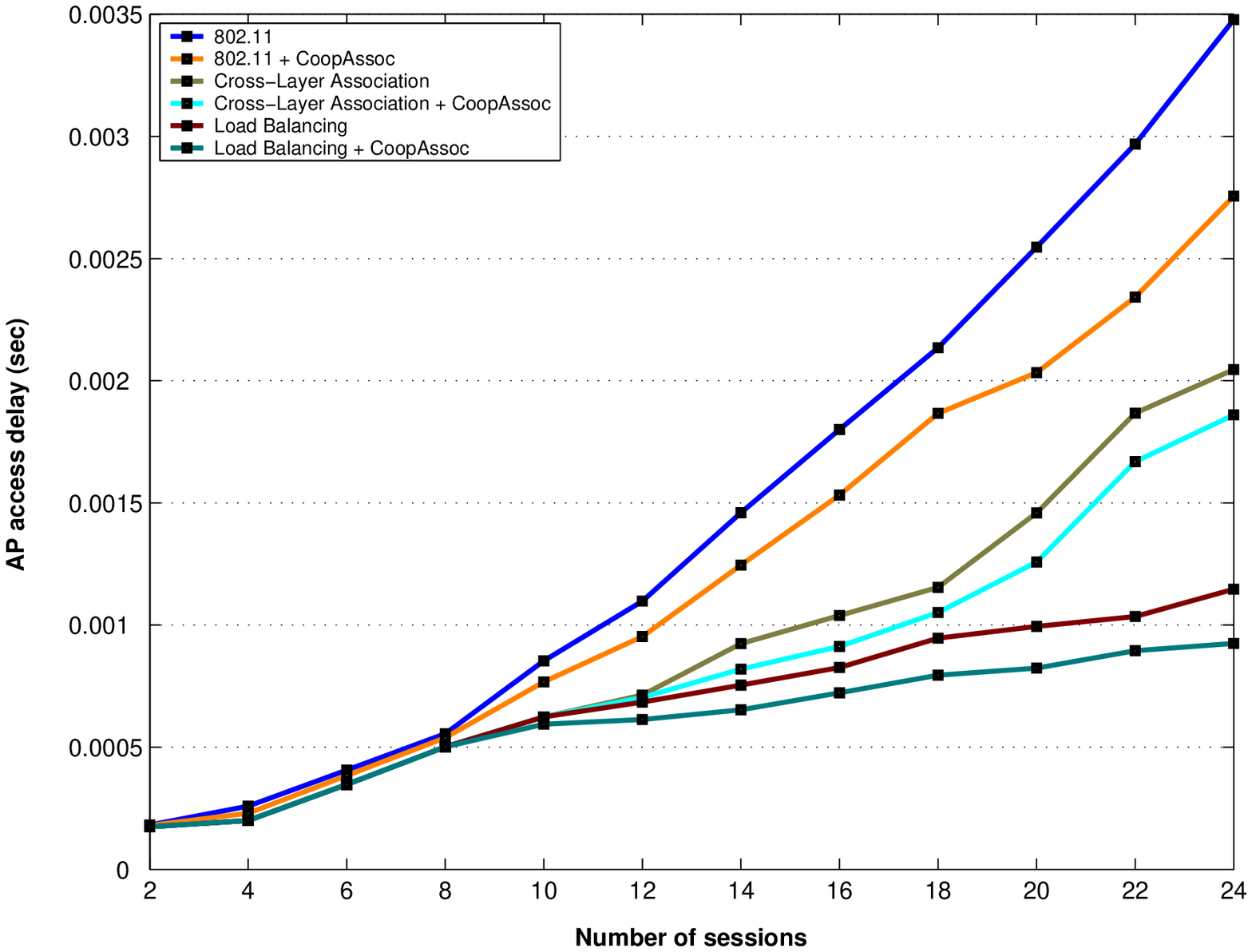}}\label{fig10b}}\vspace{-0.1in}
\caption{Average VoIP delays.}} \label{fig:fig10}\vspace{-0.2in}
\end{figure*}

First of all we measure the average local client access delay in the
network. In practice, this delay reflects the time from when the
packet is generated until it leaves the client interface. The number
of the sessions that are supported in parallel increases from 2 to
24. Figure \ref{fig10a} depicts the average VoIP client access
delay. The load balancing mechanism (enhanced with cooperation)
achieves the lower client access delays in the network. Our load
balancing mechanism minimizes the channel access delay while it
provides a "cell breathing" to the overloaded cells. The associated
STAs are optimally associated in order to maintain a balanced
network operation. Consequently, our load balancing mechanism keeps
the client access delay in low level while the traditional 802.11
operation overloads the network and the client access delay is
continually increased. In high load conditions the delay improvement
that is introduced by the load balancing mechanism is quite
impressive. \hfill

Figure \ref{fig10b} depicts the average local VoIP AP access delay
in the network. This delay is the time between the arrival of a VoIP
packet to the AP until it is ether successfully transmitted over the
wireless mesh network or dropped. It is clear that we get the same
simulation results with the client access delay. The load balancing
mechanism (enhanced with cooperation) has the best performance. In
802.11 the overloaded APs (in high load conditions) have a lot of
traffic to forward to the mesh backhaul network. The main
consequence is that the VoIP packets have to wait for a long time to
be transmitted by the APs, introducing in this way huge AP access
delays.

\begin{figure*}[ht]
\centering
\parbox{1\textwidth}{
\subfigure[Average end-to-end delay.]{\parbox{0.51\textwidth}{
\includegraphics[width=3.0in]{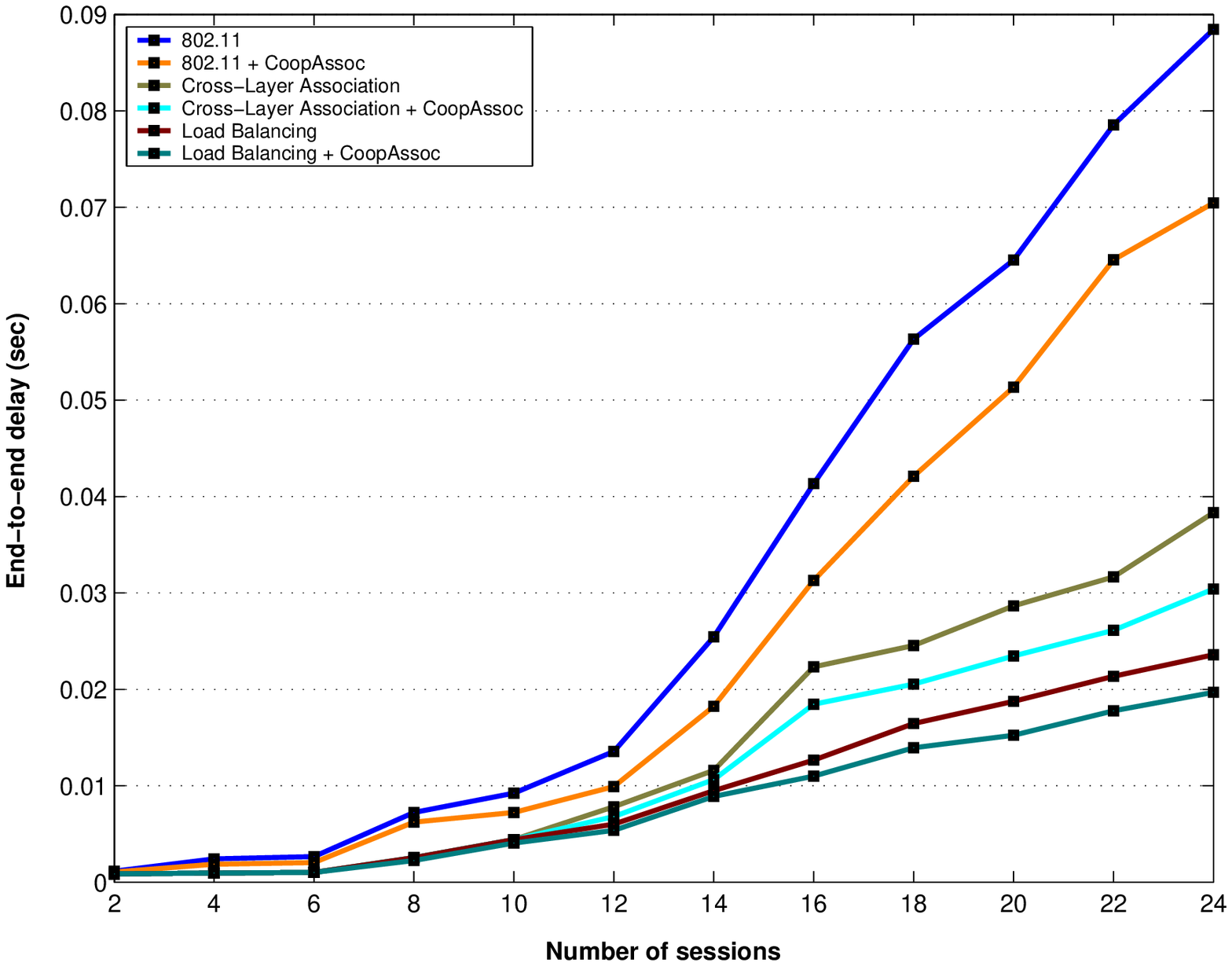}}\label{fig11a}}
\subfigure[Dropped data.]{\parbox{0.51\textwidth}{
\includegraphics[width=3.0in]{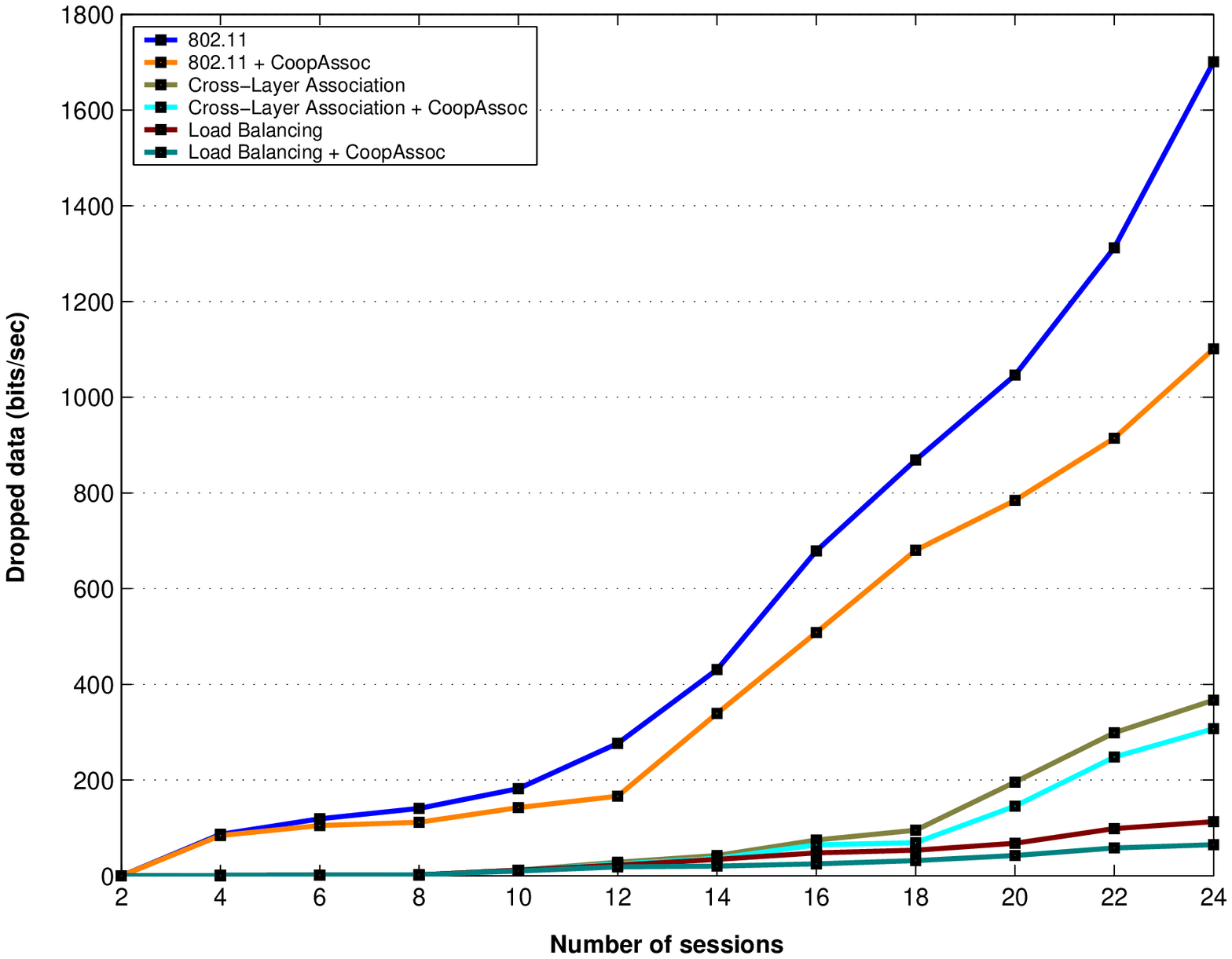}}\label{fig11b}}\vspace{-0.1in}
\caption{Average end-to-end delay and dropped data in VoIP.}}
\label{fig:fig11}\vspace{-0.2in}
\end{figure*}

In figure \ref{fig11a} we observe the average end-to-end delay in
the VoIP packet transmission. The end-to-end delay is affected by
the previous two kinds of delays that we have described in detail
and the routing delay that is introduced in the backhaul network.
The load balancing mechanism (enhanced with cooperation) achieves
lower end-to-end delays in the network. Especially in high load
network operation the delay improvement is huge. This improvement is
true due to the fast VoIP client and AP access in the network, the
effective link aware AODV routing protocol in the mesh backhaul and
the sophisticated "cell breathing" achieved by the load balancing
mechanism in overloaded cells. We argue that the most interesting
result is depicted in figure \ref{fig11a}: The pure 802.11 operation
can support at most 14 sessions in parallel while the proposed load
balancing mechanism has the capability to support 24 sessions in
parallel. Therefore we gain approximately 72,5\% network performance
improvement. The network capabilities are expanded by the use of the
sophisticated load balancing mechanism.

The last figure (figure \ref{fig11b}) depicts the dropped data
during the operation of the mesh network. Channel errors, contention
and packet collisions are the main reasons for this data dropping.
Data dropping in 802.11 is kept in high levels. Our proposed
sophisticated mechanisms decrease the data dropping and manage to
keep it low even in high load conditions.

\section{Experimental Evaluation}
In this section we present the experimental evaluation of the proposed mechanism. We evaluate our mechanisms in a wireless testbed deployed at the University of Thessaly, Greece. In the forthcoming subsections we describe the testbed deployment, we give some details about the implementation of the proposed mechanisms and we present the evaluation results.

\subsection{UTH Wireless Testbed}
The UTH Wireless Testbed has been designed to operate as a static wireless mesh network; mobility can also be facilitated by connecting laptop computers and other mobile devices and using the wireless interface(s) of those. The testbed is deployed across the 5 floors and the rooftop of the Computer and Communications campus building, in downtown Volos, Greece.

We are using ORBIT nodes \cite{Orbit} in our testbed. Each node has 2 wireline (Ethernet) interfaces and is connected to a central server, through a wired Ethernet back-haul infrastructure. One of the Ethernet interfaces is used for the network boot as well as for accessing the OS of the node and logging experiments. The second interface is connected to a Chassis manager, through which we are able to remotely power on/off the node. In conjunction to our NFS mounting strategy, this has many advantages; the main advantage is that all configurations and testbed updates are conducted centrally on the server. A simple, remote node reboot command is sufficient for all nodes to load the most-updated kernels, modules, drivers and applications. Hence, although the ORBIT nodes come with a local hard disk, we are not currently using it for OS boot. On the contrary, we have configured the nodes to retrieve their kernels, and mount their root file system directly from the central server (using the wired back-haul). This means that every researcher can maintain his/her own, independent experimental setup, including kernel, drivers, implementations, measurement logs, and every other potential component of the OS distribution. Note here that it isn't even necessary to run Linux; any operating system capable of mounting their root off NFS will work.

\begin{figure}[t]
\centering
\includegraphics[width=3.5in]{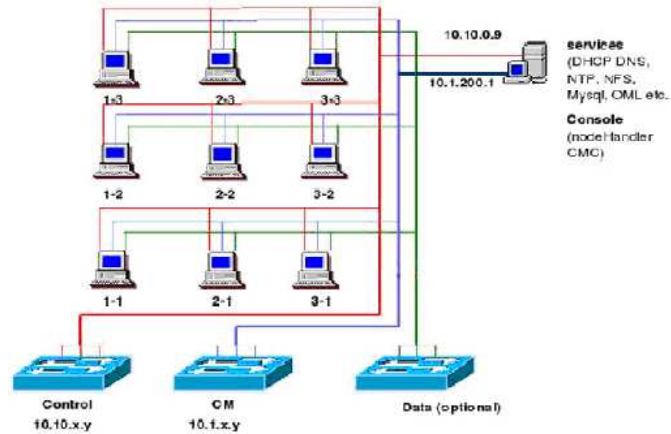}\vspace{-0.1in}
\caption{Testbed architecture.} \label{testbed1}\vspace{-0.2in}
\end{figure}

The testbed consists of wireless nodes that are based on commercial WiFi wireless cards as per the the IEEE 802.11 standard. The architecture of the testbed is similar to the ORBIT testbed in Rutgers University. The wireless nodes are connected via a wired gigabit Ethernet and the network is managed by three servers (figure \ref{testbed1}): 
\textbf{Services} - It is used to host various services including DHCP, DNS, NTP, TFTP, PXE, Frisbee, NFS, mysql, OML and Apache. We have different aliases for the management host to segregate the services that it hosts. This machine or port shall be connected with the Control port of the nodes.
\textbf{Console} - It is used to run experiments using the management software. Console is also connected with the Control port of the nodes. It may share one Ethernet port with Services. A better way is setting up a console in one machine exclusively and let it be accessible by experimenters using ssh or XDMCP.
\textbf{CMC} - It is the control and monitoring manager for all CM elements of the nodes. It is connected with the CM port of the nodes and can NOT share the same Ethernet port with Services and Console.

\begin{figure}
\centering
\includegraphics[width=2in]{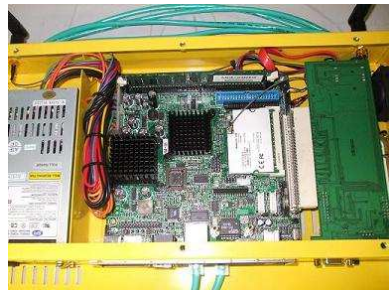}\vspace{-0.1in}
\caption{Wireless Node Architecture.} \label{testbed2}\vspace{-0.2in}
\end{figure}

Each wireless node in the testbed consists of an 1 GHz VIA C3 processor, 512 MB of RAM, 40 GB of local disk, three Ethernet ports, two 802.11 a/b/g wireless cards and a Chassis Manager (CM) to control the node. The basic architecture of a node is depicted in Fig \ref{testbed2}. The three Ethernet ports in a node are used as follows:
\textbf{Control port} - The Ethernet port between the USB ports of the node. It is a Rtl-8169 Gigabit Ethernet port, which is used to load and control the wireless node and to collect measurements.
\textbf{Data port} - The Ethernet port above the USB ports. It is a VT6102 Rhine-II 100/10baseT Ethernet port, which is used for wired data communication between the nodes.
\textbf{CM port} - The 10BaseT Ethernet port on the Chassis Manager (CM) Card, which is used to control the on/off switching of the nodes. It communicates with the management application, which controls the nodes in experimentation that is called Gridservice.

The wireless testbed is designed in a way that it can setup an infrastructure network. The backbone of the network can be either wired or wireless. For the implementation of the wired backbone among the wireless stations, we can use the data plane of the testbed (data port on each node). For the implementation of the wireless backbone, we can use in parallel both the wireless cards of the node. One card will be setup in infrastructure and one in the ad-hoc mode. In this way we can create a hybrid network that consists of a wireless backbone (cards in ad-hoc mode) that forwards the packets to the final destinations in a centralized manner (cards in infrastructure mode). In other words, we can setup an infrastructure network where the distribution system among the access points is a wireless backbone.

\subsection{Implementation and Experimental Evaluation of the Proposed Approach in the UTH Testbed}
We have used MadWifi driver \cite{MadWifi} in order to implement our scheme. MadWifi is one of the most advanced WLAN drivers available for Linux today. It is stable and has an established user-base. The driver itself is open source but depends on the proprietary Hardware Abstraction Layer (HAL) that is available in binary form only. The current stable release is v0.9.4. MadWifi has a well commented code and a large community of users and developers use it and hence it is thoroughly investigated. MadWifi is looked as the best open source driver for wireless cards for Linux as of now. It is constantly updated patched and researched by the MadWifi community.

The MadWifi driver implements most of the 802.11 MAC functionalities and therefore it is easy to modify its code in order to change parameters, or implement new features. In particular, in our protocol implementation we have changed the RSSI-based association functionality that is implemented in MadWifi and we have introduced our airtime-based association mechanism. Besides, we have extended the management frames (beacon frames) in order to carry useful information about the operational parameters of the APs in the network. The main modifications that we have introduced in the driver are:

\begin{enumerate}
\item{Every AP must measure and broadcast periodically the cumulative airtime cost in both directions (uplink and downlink): \textbf{a.} Each AP measures the transmission rate and the packet error rate (based on the transmission of the data frames) at each downlink communication and then computes the cumulative airtime cost for its downlink. As far as the computation of the packet error rate is concerned, we capture the percentage of the dropped data frames in a time window. \textbf{b.} Each associated STA captures the percentage of the dropped data frames in order to compute the packet error rate in its uplink. Then, the STAs compute their uplink airtime cost and piggy-back its value in their data frames. This is a practical way to inform the AP about the quality of the uplink communications. \textbf{c.} Each AP computes the cumulative airtime cost in both directions based on the previous measurements. \textbf{d.} Each AP periodically broadcasts the cumulative airtime cost (in its beacon frames). In order to incorporate the aforementioned value in the beacon frame we have overwritten some of its fields}
\item{Each STA that tries to find an AP to be associated with, initiates a scanning procedure. During this procedure it receives the transmitted beacon frames and captures the cumulative airtime cost of the candidate APs for association. Then, the STA decides to be associated with the AP with the minimum airtime cost.}
\end{enumerate}

\begin{figure*}[t]
\centering
\includegraphics[width=4.5in]{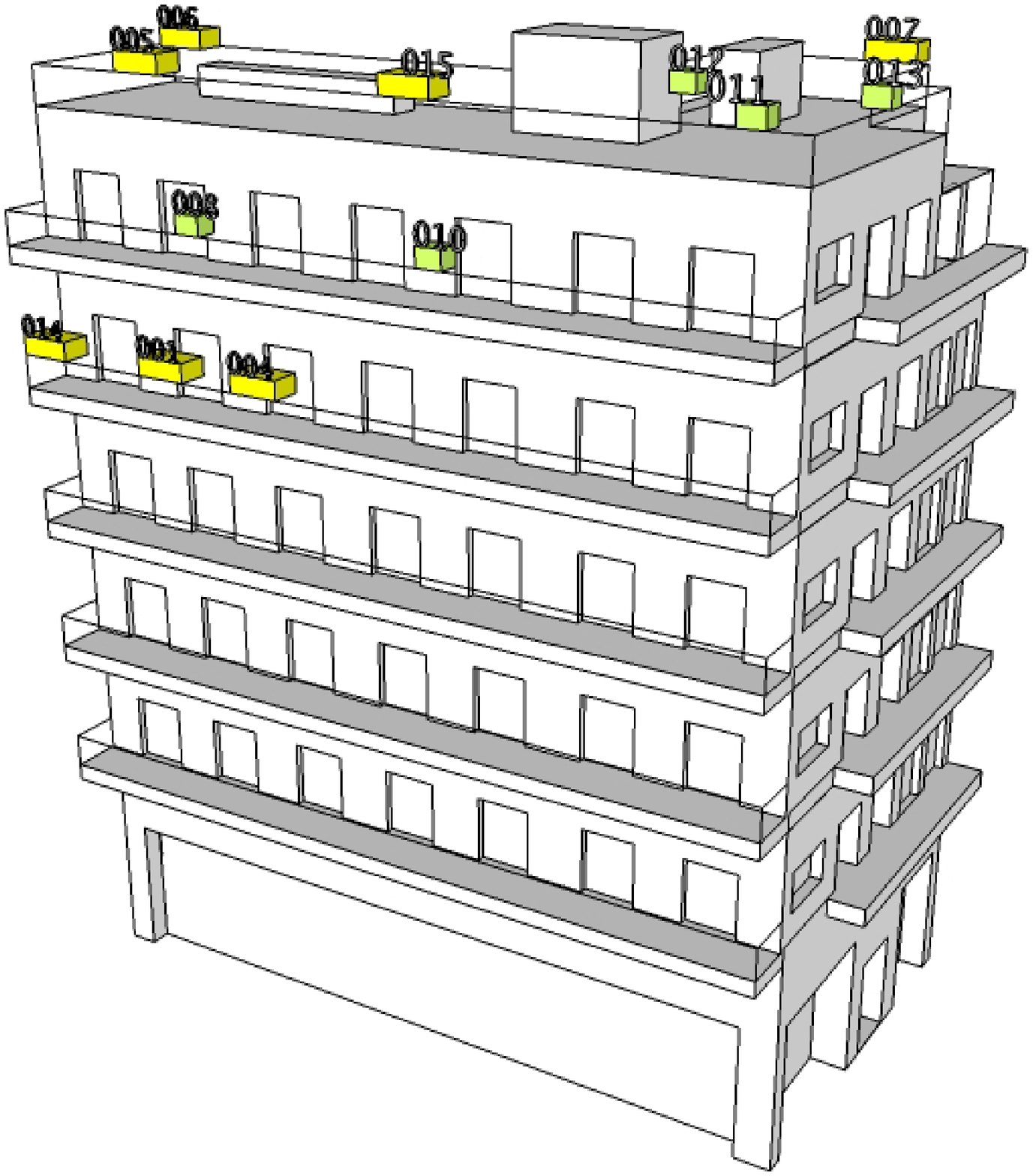}\vspace{-0.1in}
\caption{Topology of the UTH wireless testbed.} \label{top_testbed}\vspace{-0.2in}
\end{figure*}

In order to study the behavior and the scalability of the proposed association mechanism in a more realistic environment we have applied our system in the UTH wireless testbed. The topology of the testbed is depicted in figure \ref{top_testbed}. The wireless network is deployed in the 4rth, 5th and rooftop floor of the building.The testbed setup that we use in our experiments include 5 APs and 14 clients.  As far as the environmental conditions are concerned, the temperature is average and the humidity is high (these conditions affect the channel quality). The walls are supported by thick metallic skeletons, and many of them are made of brick. This degrades the signal strength on a sub-set of the links where no direct line of sight exists. The positions of the APs in the network were selected after a set of measurements and placed uniformly to ensure maximal coverage.

We must mention here that the wireless nodes are equipped with two wireless interfaces that can be used independently in our experiments. All nodes
by default set their transmission power to the maximum (20 dBm). Each client sends/receives fully saturated UDP traffic for two hours, to/from its AP(we run each experiment several times in order to get accurate results). We use the
iperf bandwidth measurement tool to generate traffic in the network and measure the performance. During each experiment, a central testbed server periodically stores information concerned to the performance of the network.

\begin{figure*}[t]
\centering
\parbox{1\textwidth}{
\subfigure[Total network throughput Vs. Number of clients.]{\parbox{0.51\textwidth}{
\includegraphics[width=3in]{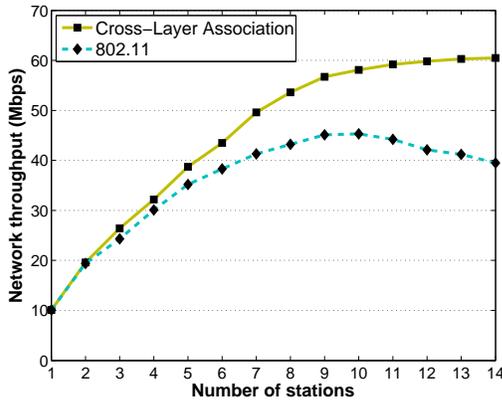}}\label{tesbed_ex1a}}
\subfigure[Average transmission delay Vs. Number of clients.]{\parbox{0.51\textwidth}{
\includegraphics[width=3in]{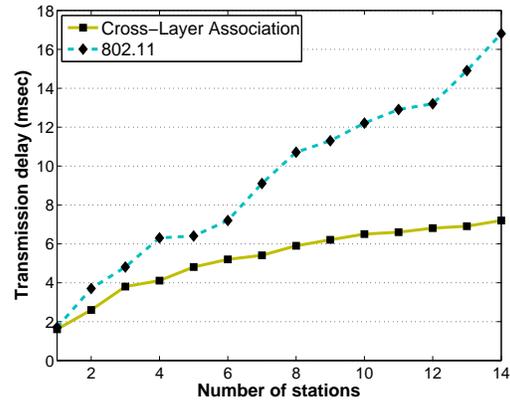}}\label{tesbed_ex1b}}\vspace{-0.1in}
\caption{Network performance in the first experiment.}}
\label{tesbed_ex1}\vspace{-0.2in}
\end{figure*}

In the first experiment we set nodes 1, 6, 7, 10, 11 as the APs in the network and the rest of the nodes act as clients. The channels that are used by the APs are selected in a way that there is no co-channel interference effects. First of all, we turn on the APs in the network and we keep the clients turned off. Then, we start turning on the clients and measure the network performance while the number of the clients in the network increases. We apply both the airtime-based association policy (modified MadWifi driver) and the pure 802.11 protocol (original MadWifi driver). We compare the network performance that is achieved under these policies. An important observation in this experiment is that the APs 1 and 11 ``attract" a lot of clients (5 clients each) in case that the RSSI-based association policy is applied. This is true due to their favorable location. An important consequence of the aforementioned situation is the overloaded performance of these two APs. APs 6, 7, 11 serve the rest of the clients in the network. It is obvious that under these operational characteristics the bandwidth is significantly wasted in the network. Figure \ref{tesbed_ex1a} depicts the performance variation of the network while the number of the clients increases. Our cross-layer association policy achieves similar performance to 802.11 when the load in the network is low. Contrarily, when the number of the clients is getting high and the load in the network increases the suboptimal operation of 802.11 drops the network performance. The airtime mechanism captures the overload effect in APs 1 and 11 (based on the transmission rate and the packet dropping), and provides load balancing by forcing some clients to be associated with the neighboring APs. The performance of the network is improved up to 52\%. Another important observation is that the total network throughput achieved by 802.11 is maximized when 10 clients are supported in the network and after that point the performance drops due to the overloading. However, our association policy keeps improving the total network throughput even if 14 clients are supported in the network. Figure \ref{tesbed_ex1b} depicts the average transmission delay when both policies are applied. Our association mechanism keeps the average transmission delay in low levels and improves the performance of 802.11 by 59\%.

\begin{figure*}[t]
\centering
\parbox{1\textwidth}{
\subfigure[Total network throughput Vs. Number of APs.]{\parbox{0.51\textwidth}{
\includegraphics[width=3in]{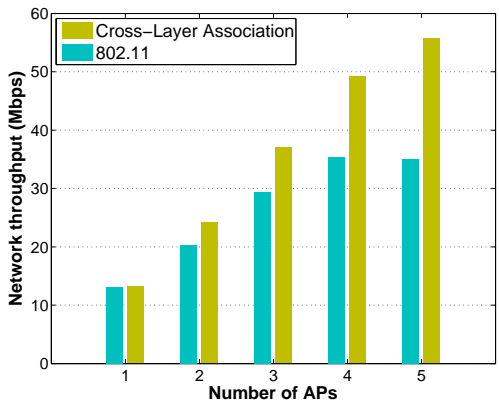}}\label{tesbed_ex2a}}
\subfigure[Average transmission delay Vs. Number of APs.]{\parbox{0.51\textwidth}{
\includegraphics[width=3in]{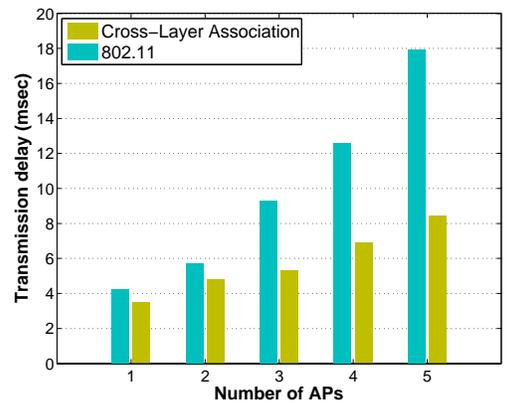}}\label{tesbed_ex2b}}\vspace{-0.1in}
\caption{Network performance in the second experiment.}}
\label{tesbed_ex2}\vspace{-0.2in}
\end{figure*}

In the second experiment we keep the same topology and the network configuration. In order to measure the scalability of the proposed association mechanism, we pick randomly one AP and 3 clients at a time and we turn them on. This process continues till all the APs and the clients are turned on in the network. Figure \ref{tesbed_ex2a} shows the total network throughput and figure \ref{tesbed_ex2b} shows the average transmission delay while the number of the APs increases. As we can see our mechanism scales much better and improves the total network throughput achieved by 802.11, by 61\%. An important observation here is that while the number of the deployed APs in the network increases, the clients act statically. In other words, the clients keep their associations with the old APs and a possible re-association is significantly delayed.  Our mechanism introduces dynamic re-associations when new light loaded APs with better channel quality are deployed in the network, providing in this way balanced network operation.

\begin{figure*}[t]
\centering
\parbox{1\textwidth}{
\subfigure[Total network throughput Vs. Number of clients.]{\parbox{0.51\textwidth}{
\includegraphics[width=3in]{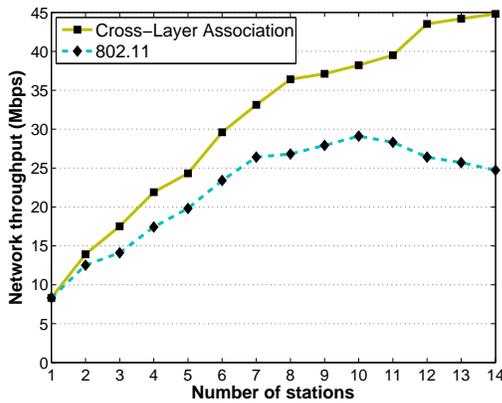}}\label{tesbed_ex3a}}
\subfigure[Average transmission delay Vs. Number of clients.]{\parbox{0.51\textwidth}{
\includegraphics[width=3in]{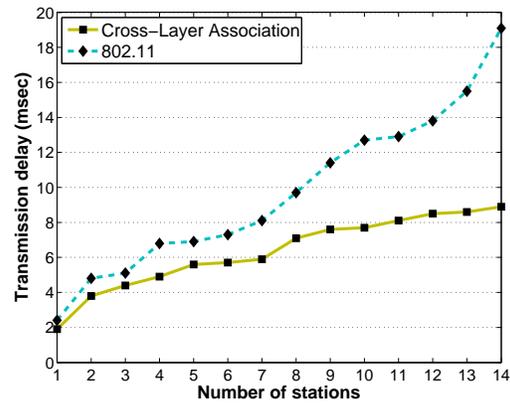}}\label{tesbed_ex3b}}\vspace{-0.1in}
\caption{Network performance in the third experiment.}}
\label{tesbed_ex3}\vspace{-0.2in}
\end{figure*}

In the third experiment we opt to introduce co-channel interference and high contention in the network. The APs 6, 10 and 11 operate on the same frequency and the rest APs operate on randomly picked frequencies. This scenario is close to real network deployments where most of the APs that are deployed operate on default frequencies or the users chose randomly channels for their APs without taking into account the interference. As we have seen from the previous experiments, the AP 11 is overloaded when the 802.11-based association procedure is applied. In the current experiment the suboptimal network performance of 802.11 is getting even worse since the AP 11 must respect the transmissions of APs 6 and 10 (must be silent when these APs are active) since they operate on the same channel. Figures \ref{tesbed_ex3a} and \ref{tesbed_ex3b} show the comparison between the cross-layer association mechanism and the 802.11. Our association policy captures the performance degradation that is introduced due to the co-channel interference and the increased contention levels, and forces the clients to be associated with the rest APs. Unfortunately, the RSSI-based association policy is not capable to capture these conditions and keeps associating the clients with the closest AP (the AP with the higher RSSI). The performance of 802.11 is improved by 84\%.

\begin{figure*}[t]
\centering
\parbox{1\textwidth}{
\subfigure[Total network throughput Vs. Number of APs.]{\parbox{0.51\textwidth}{
\includegraphics[width=3in]{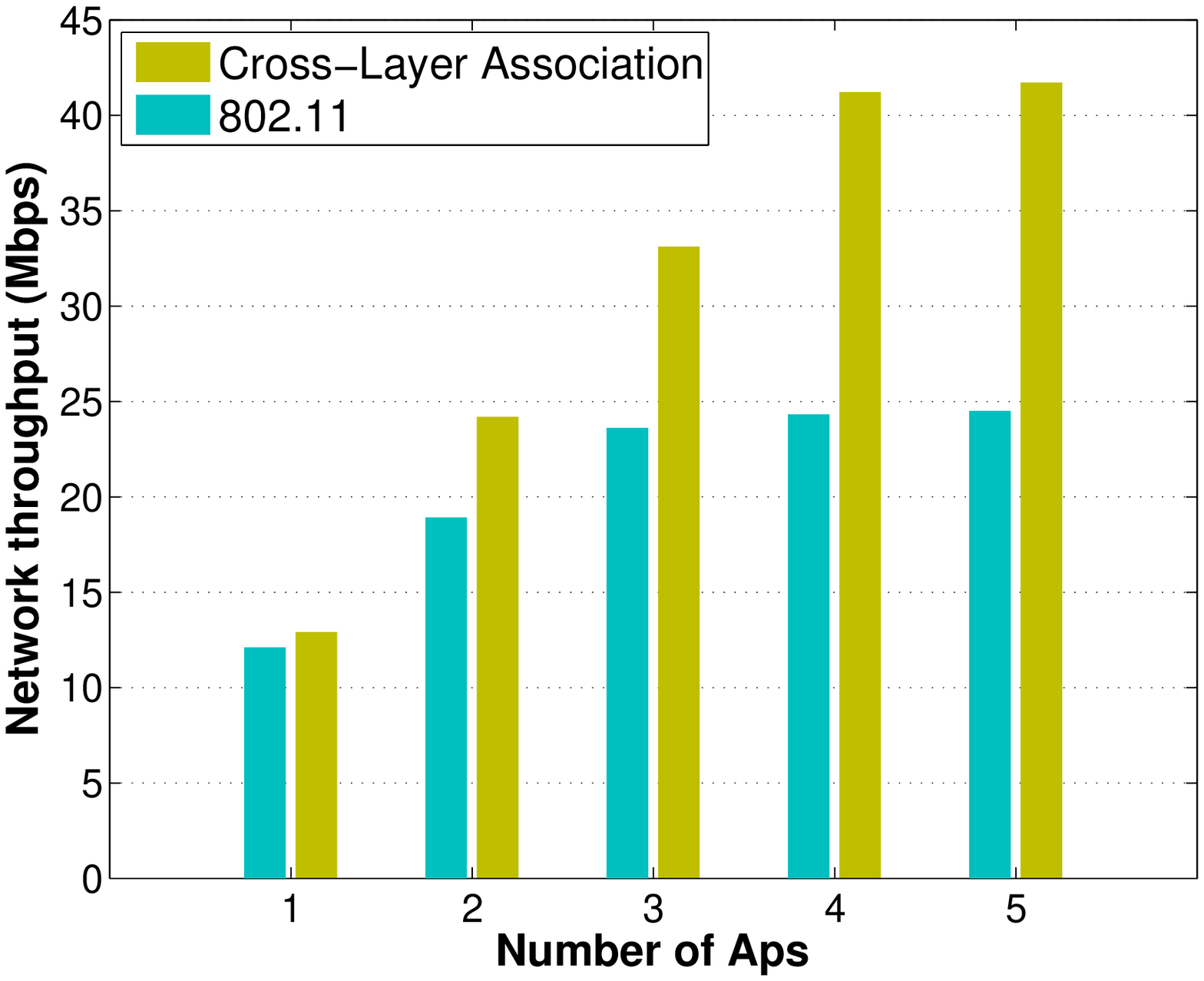}}\label{tesbed_ex4a}}
\subfigure[Average transmission delay Vs. Number of APs.]{\parbox{0.51\textwidth}{
\includegraphics[width=3in]{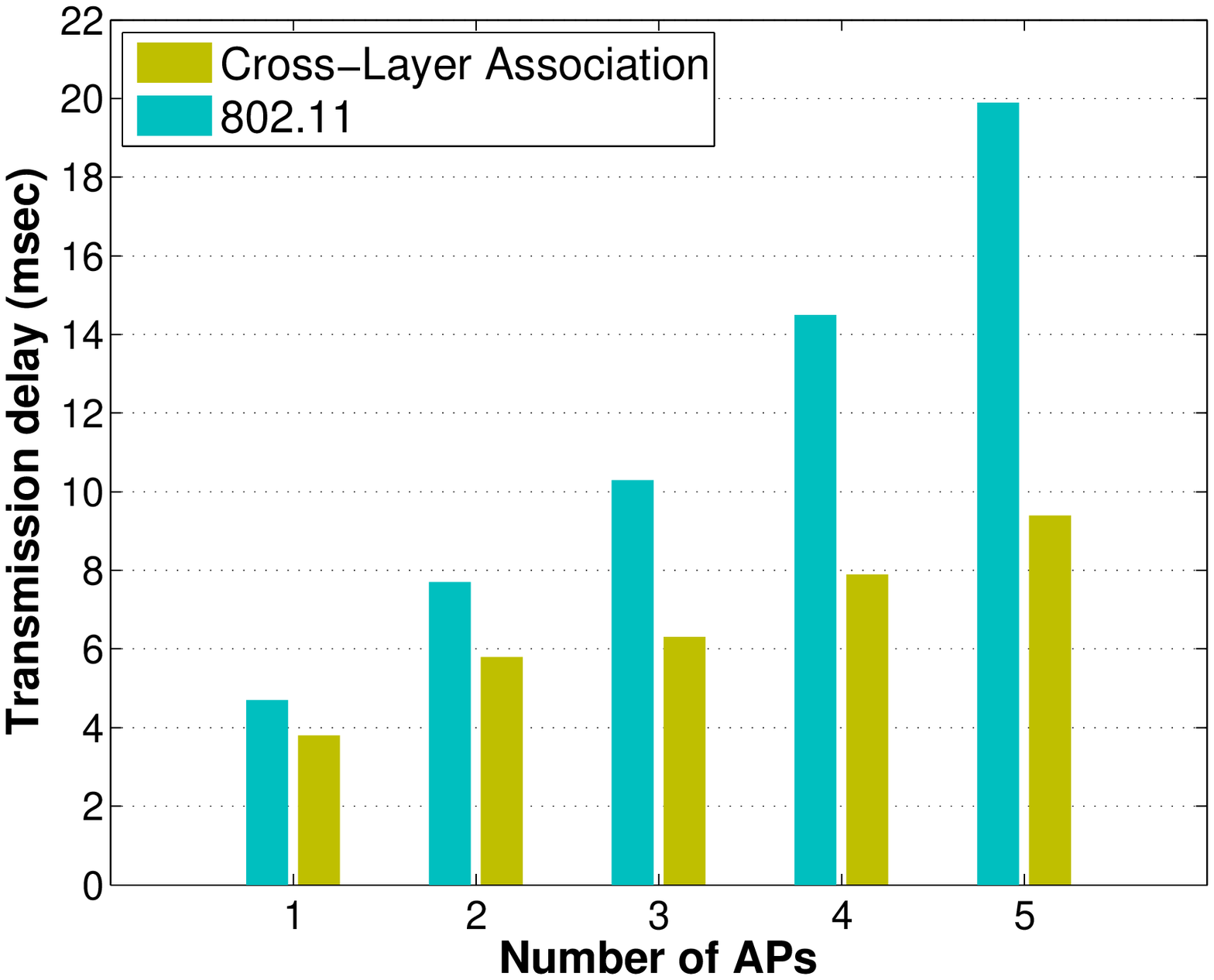}}\label{tesbed_ex4b}}\vspace{-0.1in}
\caption{Network performance in the fourth experiment.}}
\label{tesbed_ex4}\vspace{-0.2in}
\end{figure*}

In the fourth experiment we keep the previous configuration and we measure the performance of the network while the number of the deployed APs varies. In figures \ref{tesbed_ex4a} and \ref{tesbed_ex4b} we can observe that the performance improvement that is introduced by our mechanism is higher compare to the previous experiment (close to 75\%). In addition, the total network throughput is stabilized when we have more than 3 APs deployed in the network. Our mechanism applies a sophisticated association policy expanding the network capabilities and maximizing the total network throughput in presence of 5 APs in the network. In particular, the dynamic re-associations that are present under the cross-layer association mechanism provide a ``cell breathing" to the overloaded cells.

\begin{figure*}[t]
\centering
\includegraphics[width=3.2in]{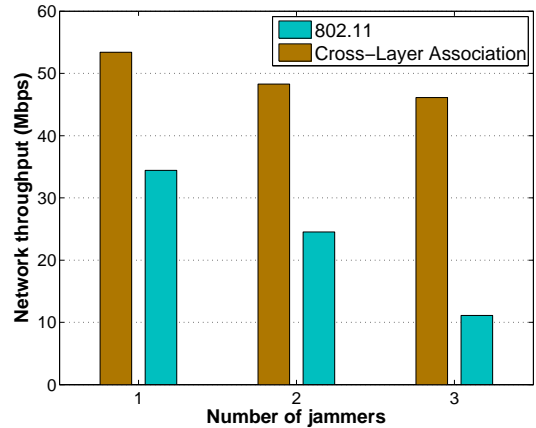}\vspace{-0.2in}
\caption{Total network throughput Vs. Number of jammers.} \label{tesbed_ex5}\vspace{-0.2in}
\end{figure*}

Continuing our experimental evaluation, we vary ``artificially" the interference level in the network. In this experiment we opt to observe the behavior of the proposed association policy when some of the APs in the network face huge amounts of interference/contention and therefore, they are enable to serve their clients. We introduce malicious clients, called jammers, that produce huge amounts of traffic in a specific channel trying to achieve denial of service in the specific part of the network. Our implementation of a constant jammer is based on a card configuration that sends broadcast packets as fast as possible. By setting the CCA threshold to 0 dBm, we force the WiFi card to ignore all
802.11 signals during carrier sensing (packets arrive at the jammer's circuitry with powers much less than 0 dBm, even if the distances between the jammer and the legitimate transceivers are very small). The jammer transmits broadcast UDP traffic. This ensures
that its packets are transmitted back-to-back and that the jammer does not wait for any
ACK messages (the back-off functionality is disabled in 802.11 for broadcast traffic). In particular, we use the clients 5, 13, 14 as jammers, that are close to the APs 1, 6 and 11. Figure \ref{tesbed_ex5} depicts the effect of the jammers in the network performance. The throughput degradation with 802.11 is very impressive (close to 73\%), especially when all the jammers are active in the network. As we have previously mentioned the jammed APs 1 and 11 serve a lot of clients. These clients face now a denial of service attack and their performance is significantly affected. Our association mechanism captures the huge amounts of interference/contention (measuring the huge packet dropping and the transmission delays) and force the jammed clients to be associated with the ``healthy" APs. In this way we limit the throughput degradation (close to 18\%).

\begin{figure*}[t]
\centering
\includegraphics[width=3.2in]{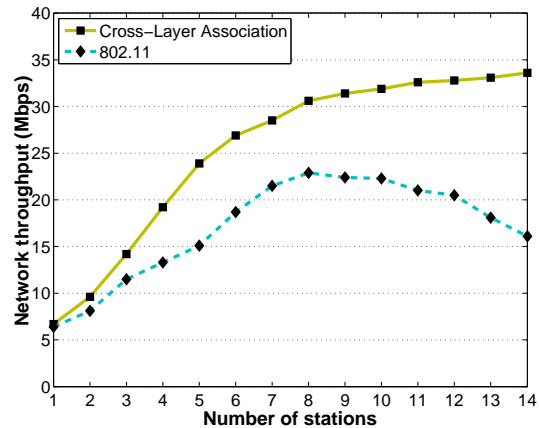}\vspace{-0.2in}
\caption{Network performance with 802.11g (Total network throughput Vs. Number of clients).} \label{tesbed_ex6}\vspace{-0.2in}
\end{figure*}

The main focus of the last experiment is to approach the operational conditions of the real 802.11 wireless network deployments. We select randomly the channels that are used by the APs in our network, without taking into account in this decision the interference from the neighboring networks. Due to the limited number of orthogonal channels in the 2.4 GHz band, the contention and the interference are high. This effect is getting even worse while most of the users set their APs in a default channel. We performed a scanning procedure in the neighborhood: 18 APs are active (outside the testbed) and 10 of them use channel 6 (default channel). Figure \ref{tesbed_ex6} depicts the total network throughput while the number of the clients varies. The cross-layer association mechanism keeps improving the network throughput by re-associating the clients that face huge interference levels with APs that are free of interference and contention, while the 802.11 is incapable to achieve high throughput values. The performance improvement that is introduced by our mechanism is quite impressive (up to 112\%).

\section{Conclusions And Future Work}
In this paper we propose a new association mechanism that introduces
cooperation between the STAs in a wireless mesh network. The
association process is executed in a cooperative manner in order to
eliminate the association/reassociation delays. Furthermore, we
propose a sophisticated load balancing scheme that guarantees a
balanced network operation. The heuristic algorithms jointly balance
the MAC/Routing communication load. Our main contributions in the
current research field are: 1) A new association framework in wireless mesh networks that introduces cooperation between the STAs, 2) A cross-layer load balancing mechanism that can be applied in overloaded cells in the network and 3) Extensive simulations and testbed experiments where we support QoS sensitive applications. We jointly implement our mechanisms and we measure the performance improvement that is achieved.
Our future directions include extending the
cooperative concept and jointly apply the proposed mechanisms in combination with power
control and channel allocation policies in order to build
a complete
cross-layer resource management system in wireless mesh networks.



%


\end{document}